# Building music with Lego bricks and Raspberry Pi

Ana M. Barbancho[1] · Lorenzo J. Tardón[1]  · Isabel Barbancho[1]

## Abstract

In this paper, a system to build music in an intuitive and accessible way, with Lego bricks, is presented. The system makes use of the new powerful and cheap possibilities that technology offers for making old things in a new way. The Raspberry Pi is used to control the system and run the necessary algorithms, customized Lego bricks are used for building melodies, custom electronic designs, software pieces and 3D printed parts complete the items employed. The system designed is modular, it allows creating melodies with chords and percussion or just melodies or perform as a beatbox or a melody box. The main interaction with the system is made using Lego-type building blocks. Tests have demonstrated its versatility and ease of use, as well as its usefulness in music learning for both children and adults.

**Keywords** Beatbox · Melody box · Lego · Raspberry Pi · MIDI · Music composition

## 1 Introduction

Music and technology are part of the everyday life for many people in the whole world. There is a growing number of original ways of creating, interacting, and playing with music using different possibilities technology can offer. For several years now, new systems appear constantly in which the musical experience is more and more immersive and does not rely on physical instruments, but generic commercial or specific systems for motion detection [15, 16, 24, 34, 40] or the interaction with elements of a physical desk (pencil, pen, table, etc.) [12] or other interaction means. In addition, more and more studies

---

Ana M. Barbancho, Lorenzo J. Tardón and Isabel Barbancho are authors contributed equally to this work

✉ Lorenzo J. Tardón
ltg@uma.es

Ana M. Barbancho
abp@uma.es

Isabel Barbancho
ibp@uma.es

[1] ATIC Research Group, ETSI Telecomunicación, Universidad de Málaga, Campus Teatinos, 29071 Málaga, Spain

are emerging to understand music in relation to the way people interact with it [11, 13]. Another good example of the work carried out within this framework is the H2020 project #MusicBricks [37] that presents many and diverse interfaces of interaction and music creation oriented to both entertainment and music professionals.

Music composition is commonly seen as a complex task that can only be accomplished by experienced, well-trained individuals or by specialized, automatic systems [14, 20, 25, 32, 33, 41]. This fact makes this creative activity to be beyond the reach of most people.

However, music composition systems for children are beginning to emerge as a creative element in their education [1], starting in kindergarten [18]. These systems use Lego pieces as a manipulative element, but the generation of sounds and the actual composition must be done on a computer [2]. There are other Lego-based music learning systems that use piano roll-like notation [21, 26, 28].

In this context, we present a new way of music creation using modified actual Lego-Duplo building blocks to enable most people to create music melodies with a rhythmic base and even chord accompaniment by building with bricks. The target audience is especially children, but it can also be used with elder people to train in therapeutic psychomotricity. The system consists of three main parts: one to synchronize (Synchronism Box), another one to set the rhythm (beatbox), and a third one to create the melody (melody box). Each part of the system is controlled using a Raspberry Pi (RPI) [31]. Note that our approach does not aim to give aids for the composition of musical melodies or automatically create music [14, 20, 25, 32, 33, 41] but to provide a different playful way to approach basic concepts of music and music creation.

The paper outline is as follows: in Sect. 2, the block diagram of the whole system and the description of the three main subsystems and the communication between subsystems are presented. Specifically: in subsection 2.1 the data communication protocol between the three subsystems is presented; subsection 2.2 describes the synchronism subsystem used to synchronize signals and processes for the whole system; next, in subsection 2.3 the beatbox subsystems that create the rhythmic base of percussion to accompany the melody is presented; the description of the subsystems ends with subsection 2.4, in which details of the melody subsystem are exposed, this part is responsible for creating the melody and accompanying chords. Section 3 displays several usage examples, results, and a discussion on the evaluations carried out by external users. Finally, the last section presents the conclusions drawn from this work.

## 2 System design

Figure 1 shows a block diagram of the whole system. This is an interactive music demonstrator that allows creating a melody with chords and percussion easily and intuitively using Lego-type building blocks. The system can work as just a beatbox or just a melody box, by simply removing the melody box or the beatbox, respectively, as the diagram in Fig. 1 illustrates.

The complete block diagram including the systems, data and control lines and external devices: sound card, computer and speaker of the devised system is presented in Fig. 2. In this figure, it can be observed that three main subsystems are designed: synchronism box, beatbox and melody box. These parts are interconnected by MIDI cables using standard MIDI protocol. Three separate Raspberry Pi's are used to control and process all the signals and ensure modularity. Finally, in order to convert the standard MIDI signal to sound, an external sound card connected to a computer running a VST plug-in is needed. Raspberry Pi's are used to control all the parts of the system separately. Its versatility, processing capacity, and the different available physical ports included make it possible. In Fig. 3, the input/output ports of the Raspberry Pi are shown; these include 26 GPIO (General Purpose Input Output) pins. These pins can be used as input, output, or both and are controllable by the user at runtime. Among the GPIO pins, 5 of them allow communication through the SPI (Serial Peripheral Interface) protocol [7] (note that SPI needs to be enabled), 2 pins that allow communication through the I2C (Inter-Integrated Circuit) protocol [27] and 2 others intended for UART (Universal Asynchronous Receiver-Transmitter) connection [30] for serial port communication. Note that, it must be taken into account that the GPIO pins offer a voltage of 3.3 V and are not tolerant to voltages of 5 V; this fact is relevant for the system design.

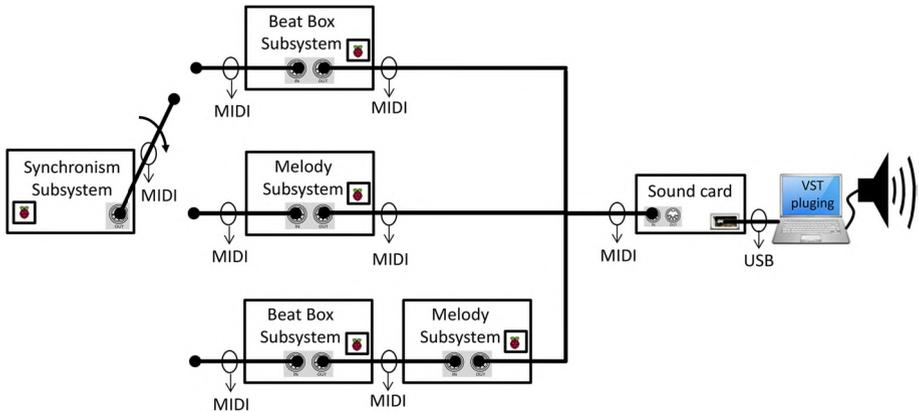

**Fig. 1** Block diagram of the system for building music with Lego bricks using Raspberry Pi

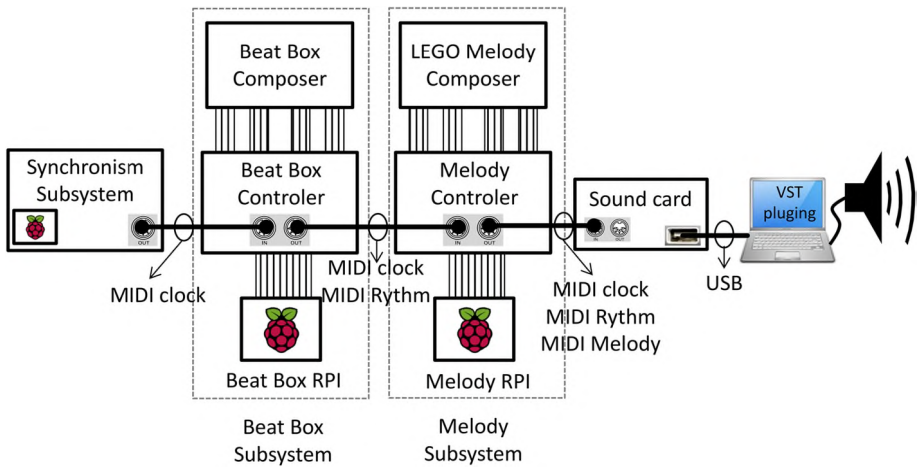

**Fig. 2** Detailed block diagram of the system for building music with Lego bricks using Raspberry Pi

**Fig. 3** Raspberry Pi IO pins

| | Pi Model B/B+ | | |
|---|---|---|---|
| 3V3 Power | 1 | 2 | 5V Power |
| GPIO2 SDA1 I2C | 3 | 4 | 5V Power |
| GPIO3 SCL1 I2C | 5 | 6 | Ground |
| GPIO4 | 7 | 8 | GPIO14 UART0_TXD |
| Ground | 9 | 10 | GPIO15 UARTA0_RXD |
| GPIO17 | 11 | 12 | GPIO18 PCM_CLK |
| GPIO27 | 13 | 14 | Ground |
| GPIO22 | 15 | 16 | GPIO23 |
| 3V3 Power | 17 | 18 | GPIO24 |
| GPIO10 SPI0_MOSI | 19 | 20 | Ground |
| GPIO9 SPI0_MISO | 21 | 22 | GPIO25 |
| GPIO11 SPI0_SCLK | 23 | 24 | GPIO8 SPI0_CE0_N |
| Ground | 25 | 26 | GPIO7 SPI0_CE1_N |
| ID_SD I2V ID EEPROM | 27 | 28 | ID_SC I2C ID EEPROM |
| GPIO5 | 29 | 30 | Ground |
| GPIO6 | 31 | 32 | GPIO12 |
| GPIO13 | 33 | 34 | Ground |
| GPIO19 | 35 | 36 | GPIO16 |
| CPIO26 | 37 | 38 | GPIO20 |
| Ground | 39 | 40 | GPIO21 |

### 2.1 Data communication

Data communication between the three subsystems shown in Fig. 1 and Fig. 2 is carried out by using the standard MIDI protocol. This choice has inherent hardware implications, specifically: the use of 5-pin DIN connectors and the voltage levels employed by MIDI signals, which require voltages between 3.5 V and 5 V to represent the high logic level. Recall that Raspberry Pi works at 3.3 V maximum. In order to solve this issue, additional hardware was designed for the MIDI out and MIDI in connections. The schematics for both the MIDI in and MIDI out connections are shown in Fig. 4 and Fig. 5, respectively [23].

The hardware created for MIDI out is shown in Fig. 4. This design allows to convert the 3.3 V output of pin UART-Tx of Raspberry Pi (RPI) to a 5 V output that can be driven to the 5-pin DIN MIDI connector.

The hardware for MIDI in is shown in Fig. 5. Observe that this interface needs a more elaborated design, and additional hardware since it is necessary to isolate the input signal, which ranges between 0 and 5 V, from the UART-Rx input of RPI, which handles up to 3.3 V. This is achieved by the utilization of an opto-isolator, as shown in Fig. 5 [23].

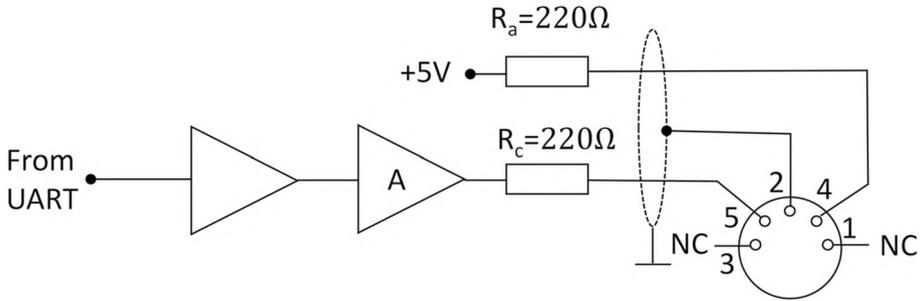

**Fig. 4** MIDI out additional hardware [23]. Voltage level adaptation

## 2.2 Synchronism subsystem

The first subsystem shown in Fig. 2 is devoted to provide with the synchronization signal for the whole system. This subsystem creates a MIDI clock signal that will be sent through the whole system to allow for synchronized tempo for music composition.

In this subsystem, beats per minute parameter (BPM) is set, together with the time signature, and the maximum number of quarter notes the melody and the rhythm signal will have before the end of a cycle.

In Fig. 6, the MIDI clock signal is shown conceptually. In this case, the maximum amount of quarter notes considered is 16.

For each quarter-note, a Continue signal is sent. During each quarter-note period, 24 Timing Clock signals are sent. For every 16 quarter-notes, a Start signal is sent. The time between each quarter note is set by means of the selected BPM, so that the duration of a quarter note is $= 60/BPM\ sec$. The use of the Continue MIDI message is standard in sequencers and it is needed in order to decide with quarter note, among the 16 in the board, must be read. The maximum $BPM$ value is limited to 208; this limitation is due to the processing time of the RPI.

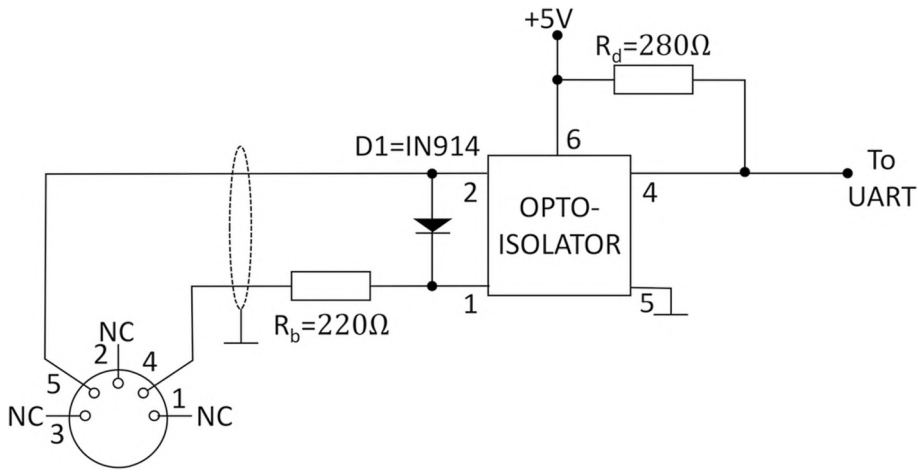

**Fig. 5** MIDI in additional hardware [23]. Voltage level adaptation

**Fig. 6** Illustration of the clock signal in a MIDI sequence

| MIDI-MESSAGE (Hexadecimal) | EVENT |
|---|---|
| FA | Start |
| F8 | Timing Clock |
| F8 | Timing Clock |
| ⋮ | ⋮ |
| F8 | Timing Clock |
| FB | Continue |
| F8 | Timing Clock |
| ⋮ | ⋮ |
| F8 | Timing Clock |
| FB | Continue |
| ⋮ | ⋮ |
| FB | Continue |

(24 times, 24 times, 16 times)

The clock signal is sent through GPIO14 (UART0_TXD) UART (Universal Asynchronous Receiver/Transmitter) of the RPI (see Fig. 3). This is an asynchronous serial communication protocol.

### 2.3 Beatbox subsystem

After the Synchronism subsystem, the beatbox subsystem is described. The aim of this subsystem is to create the rhythmic base of percussion to accompany the melody. Note that this subsystem can be removed from the chain to let the melody subsystem alone with the synchronization one, or it can be used without the melody subsystem as well (Fig. 1). In order to achieve this behaviour, the output of this box, MIDI Rhythm, must be added to the MIDI clock.

It must be taken into account that, in General MIDI standard streams, channel 10 is reserved for percussion instruments [9]: each MIDI note in this channel corresponds to a single note of a percussion instrument. In the system presented, four different percussion instruments, with total duration corresponding to 16 quarter-notes, are considered.

As it is shown in Fig. 2, the beatbox is divided into tree blocks:

- Beatbox composer: this is the hardware structure for building the rhythmic pattern in a visual and intuitive way.
- Beatbox controller: this part includes a MIDI-out port, and a MIDI-in port. These pieces, described in Sect. 2.1, and Figs. 4 and 5, are designed to allow the connection of the Lego beatbox hardware to the RPI.
- Beatbox RPI. The RPI is connected to the beatbox controller by one 40-flat cable grid.

These three blocks are described in deeper detail next.

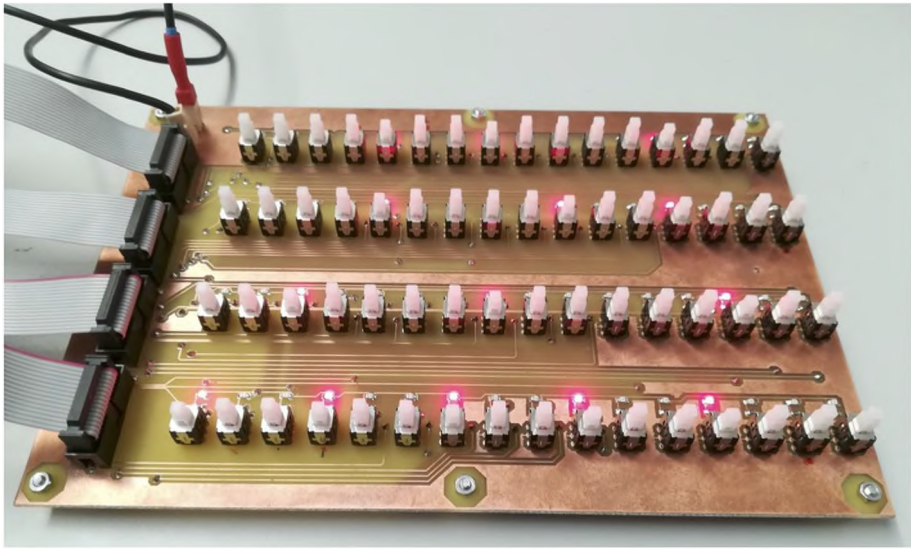

**Fig. 7** Hardware structure for the rhythmic pattern builder

### 2.3.1 Beatbox composer

The structure for building the rhythmic pattern is shown in Fig. 7. It is a printed circuit board of dimensions $16 \times 23.5 cm$, on which the components are soldered. The logical mapping of the hardware shown in Fig. 7 into its musical counterpart is exposed in Fig. 8.

As shown in Fig. 7, the composer includes a push-button with light for each of the 16 quarter-note periods of the 4 percussion instruments selected. This leads to $16 \times 4$ push-buttons. So, the output of the beatbox composer is formed of $16 \times 4$ lines. When a button is pressed the line connected to ground and the output is 0 V, otherwise, it is 3.3 V. This output is connected to a controller to carry all data to the RPI.

### 2.3.2 Beatbox controller

Each instrument in the beatbox composer allows for 16 quarter-notes. Since the RPI does not have 64 entries, several multiplexers are needed to adapt the input to the controller. Specifically SN74LS151N [38] multiplexers are selected for the 8 to 1 multiplexing and SN74LS157 [39] multiplexer for the four 2 to 1 multiplexing. Figure 9 shows the circuit diagram for each percussion instrument. The control signals A, B and C are common to the two multiplexers 8 to 1. The control signal D that reaches the multiplexer from 2 to 1 determines if the definitive output of the channel belongs to the first 8 quarter-notes or the last 8 ones. The scheme in Fig. 9 is repeated for each of the 4 percussion instruments.

| | 1 | 2 | 3 | 4 | 5 | 6 | 7 | 8 | 9 | 10 | 11 | 12 | 13 | 14 | 15 | 16 |
|---|---|---|---|---|---|---|---|---|---|---|---|---|---|---|---|---|
| Charlie 2 | | | ■ | | | | | | | | | | | | | |
| Charlie 1 | ■ | ■ | | | ■ | | ■ | | | | ■ | | | ■ | | ■ |
| Box | | | | ■ | | | | | | | | ■ | | | | |
| Bass Drum | ■ | | | | | | | ■ | | | ■ | | | | | ■ |

**Fig. 8** Logical mapping of the rhythmic pattern builder

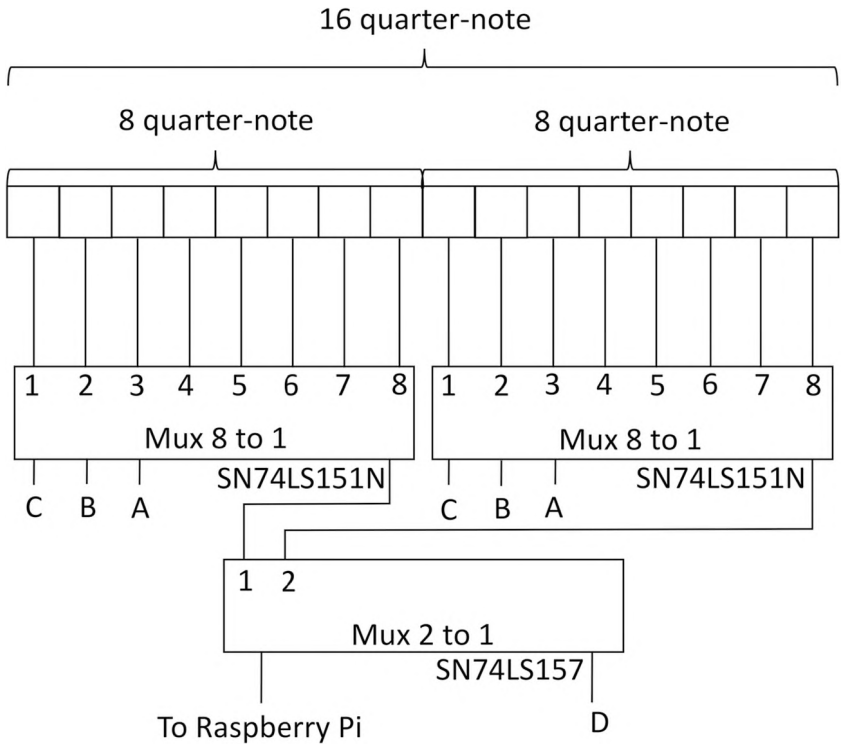

**Fig. 9** Scheme of the circuit designed to adapt each percussion instrument to the controller input

### 2.3.3 Beatbox RPI

The Beatbox RPI receives the MIDI clock through GPIO15 (UART0_RXD) (see the physical ports of the RPI drawn in Fig. 3).

The function of the beatbox RPI is to turn the rhythmic base of the percussion into a MIDI stream, and combine it with the MIDI clock signal.

The combined structure of the MIDI sequence of the clock signal and rhythm signal is represented in Fig. 10. The rhythmic base is included after the Start signal and the first 24 Timing Clock events. The MIDI sequence for the rhythmic pattern for each quarter-note period includes Mode change to channel 10 (percussion instruments) [23], All Notes Off (to turn off the previous percussion instruments), and, finally, the active notes. Figure 10 presents an illustration in which in a certain quarter-note period all the four instruments (bass drum, box, charlie 1 and charlie 2) are active.

**Fig. 10** Illustration of a MIDI sequence with clock and rhythm signals

| MIDI-MESSAGE (Hexadecimal) | EVENT |
|---|---|
| FA | Start |
| F8 | Timing Clock |
| F8 | Timing Clock |
| ⋮ | ⋮ |
| F8 | Timing Clock |
| B9 | Chan 10 Mode Change |
| 7B | All Notes Off |
| 00 | All Notes Off |
| 99 | Chan 10 Note On |
| 24 | Bass drum |
| 99 | Chan 10 Note On |
| 26 | Box |
| 99 | Chan 10 Note On |
| 31 | Charlie 1 |
| 99 | Chan 10 Note On |
| 2E | Charlie 2 |
| FB | Continue |
| F8 | Timing Clock |
| ⋮ | ⋮ |
| F8 | Timing Clock |
| ⋮ | ⋮ |
| FB | Continue |
| ⋮ | ⋮ |
| FB | Continue |

(Brackets indicate: first F8 group = 24 times; Chan 10 section through Charlie 2 = Rhythmic; second F8 group = 24 times; overall Continue block = 15 times)

### 2.4 Melody subsystem

This subsystem is aimed to create the melody and accompanying chords. So, the output of this subsystem will be a MIDI stream composed of the melody, accompanying chords, rhythm, and clock. The MIDI channel 1, configured as polyphonic, is used to send the melody, and the accompanying chords to the VST plug-in.

The Melody subsystem is composed of tree blocks (see Fig. 2):

- Lego melody composer: this is the structure created for building the melody in a visual and simple way.

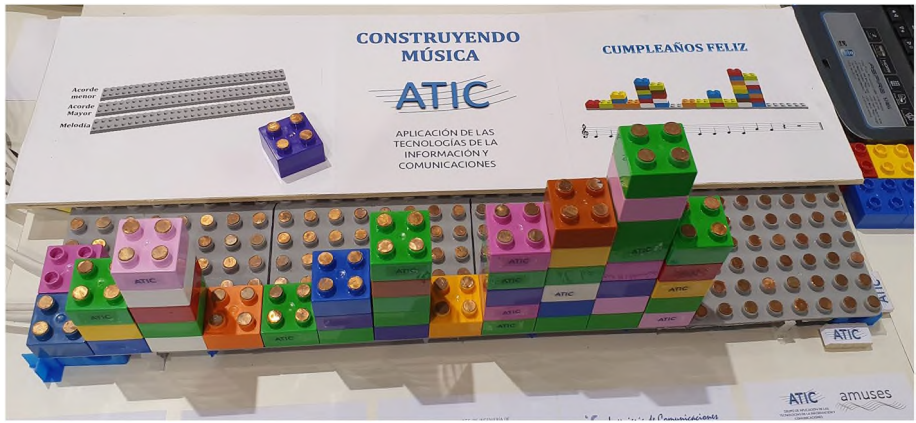

**Fig. 11** Structure for building music: melody and accompanying chords

- Melody controller: this block includes a MIDI-out port, and a MIDI-in port with the adaptation hardware presented in Figs. 4 and 5 and the necessary hardware that allows to connecting the Lego melody composer to the RPI.
- Melody RPI. The RPI runs the software to create the music from the constructions built by using the Lego melody composer scheme. It is connected to the melody controller by a 40-line flat cable.

### 2.4.1 Lego melody composer

Figure 11 shows and example of construction of music, including melody and accompanying chords.

Figure 12 shows the transcription into a music score of the melody and accompaniment built in Fig. 11.

The base for the construction of melodies, Fig. 11, consists of a Lego Duplo board with dimensions $9.5 \times 51 cm$, containing $4 \times 16$ pairs of studs, although just $3 \times 16$ pairs of studs are employed in our design. The three construction rows of pairs of studs used in our design are clearly shown in this figure; each of them has a different function:

- The first row is used to build the melody. Figure 13 shows how the pitch of each quarter-note is encoded by building pitch-towers with Lego-type pieces. It can be observed that the note range considered is 11: from $C4$ to $B4$. The lowest note ($C4$) is encoded using one Lego piece. Then, the pitch is raised half a tone for each new piece stacked. If no piece is placed, a quarter-note rest is interpreted.

Lego pieces need to be modified to build a melody so that the RPI can identify the pitch desired at each quarter-note period. The pieces are modified as shown in Fig. 14;

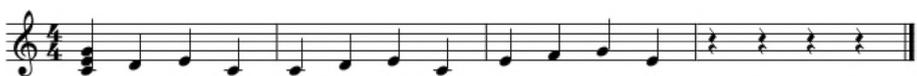

**Fig. 12** Score corresponding to the music excerpt built in Fig. 11

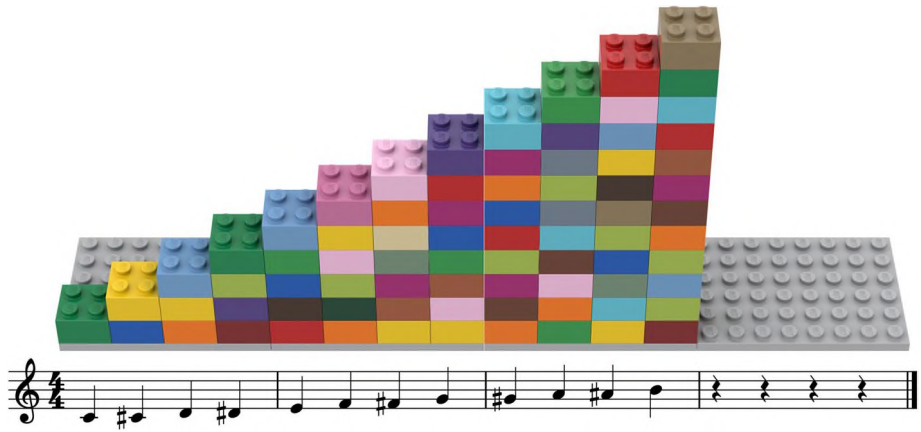

**Fig. 13** Illustration of the construction of a melody by using pitch-towers, and corresponding score

each piece contains a $50K\Omega$ resistor connected in such a way that stacking pieces means to connect resistors in parallel. Therefore, the equivalent resistance of certain number *n* of stacked pieces will be determined by equation (1):

$$Z_2(K\Omega) = \frac{1}{N\frac{1}{50K\Omega}} \qquad (1)$$

Then, the stacked pieces are connected to a voltage divisor structure as shown in Fig. 15, where $Z_1 = 10K\Omega$. Thus, the voltage, $V_{out}$, for each quarter-note period depends on the equivalent resistance of the stacked pieces according to Eq. (2):

$$V_{out} = V_{in}\frac{Z_2}{Z_1 + Z_2} \qquad (2)$$

- The second and third rows are used to determine if the melody note has a minor or major chord as accompaniment, respectively. If both pieces are connected, the system always decides major chord. Lego pieces have been customized for chord selection as shown in Fig. 16. These pieces are different to the ones used to build the melody since their only function is to determine whether a chord is enabled or not; so they are built to simply create a short circuit when placed on the board.

**Fig. 14** Detail of modified Lego pieces employed for melody construction

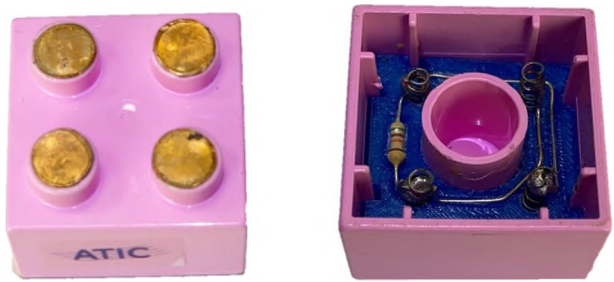

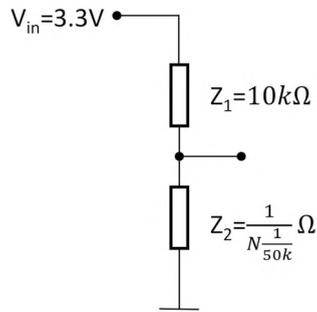

**Fig. 15** Voltage divisor structure employed to define the pitch of each quarter-note period

The Lego melody composer is connected to the melody controller with four 16-line flat cables; the additional row with 16 × 1 pairs of Lego studs provide capacity for further extensions.

### 2.4.2 Melody controller

The Raspberry Pi must convert the sequence of stacked Lego pieces in the first row of the Lego melody composer into the corresponding musical sounds to be inserted into the MIDI stream. However, recall that the Raspberry Pi only has digital inputs so, in order to interpret the inputs from the composer, an analogue to digital converter is used. Specifically, the MCP3008 has been chosen [22]. The MCP3008 is a low-cost 8-channel 10-bit analogue to digital converter that performs the conversion presented in Eq. (3):

$$DOC = \frac{1024 V_{out}}{3.3V} \qquad (3)$$

where

- DOC stands for the Digital Output Code of the converter.
- $V_{out}$ is the voltage that depends on the equivalent resistance of the stacked pieces, defined in Eq. (2).

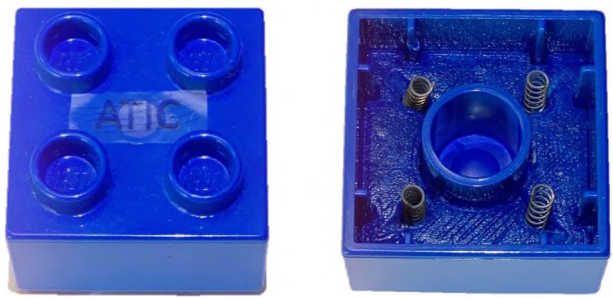

**Fig. 16** Detail of customized LEGO pieces for the selection of accompanying chords

Note that, due to the tolerance of the resistor (5% for Lego pieces and 10% for reference resistors), the Digital Output Code from Eq. (3) must be considered a value in the interval corresponding to a certain note $i$, within the range of $DOC$ values for each note in the system. Due to the fact that the notes range from C4 to B4, the note identifier, $i$, is in the interval [1, 11]. Thus, note $i$ will be decided if $DOC_i$ is in the interval expressed by equation (4):

$$DOC_i - \frac{DOC_i - DOC_{(i-1)}}{2} < DOC < DOC_i + \frac{DOC_{(i+1)} - DOC_i}{2} \quad (4)$$

where $DOC_i$ is the centre of each decision interval obtained assuming ideal resistor values.

Since the maximum length of the melody is 16 quarter-note periods, two MCP3008 are employed, and their outputs are driven to a multiplexer. The Raspberry Pi will control which signal goes to the line MELODY and convert the $DOC$ into the corresponding MIDI channel. Figure 17 shows the overall block diagram: the control of the two MCP3008 chips is done by using one set of GPIO outputs, and only one pin is used to control of the SN74LS151N multiplexer [38].

The second and third rows of the Lego melody composer have simpler operation since the connection of each Lego piece simply changes the voltage level form $3.3V$ to $0V$; consequently, only two levels of the AD converter are needed.

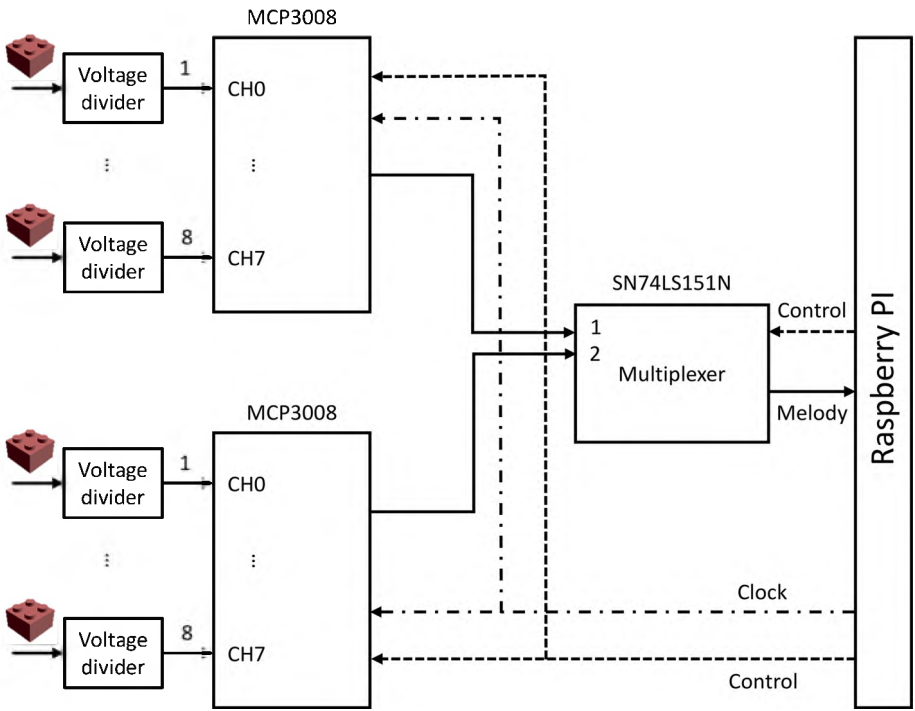

**Fig. 17** Overall block diagram of the communication architecture between the melody construction hardware and the RPI

### 2.4.3 Melody RPI

The melody RPI uses the information received from the composer through the controller to create and insert the corresponding MIDI events into the MIDI stream. The structure of the MIDI sequence including the clock, rhythm, melody, and chord is represented in Fig. 18.

The melody and accompanying chords are inserted after the rhythm signal. The MIDI sequence for the melody and accompaniment for each quarter-note includes Mode change to channel 1 piano, All Notes Off, to turn off the previous notes, and, finally, the active notes. Figure 18 shows the case in which a certain quarter-note period contains both melody and accompanying chord (4 notes). Note that, thanks to the use of MIDI protocol, the sound of different instruments [23] can be simulated by simply changing the channel number.

The system as it is built does not allow for immediate expansion. In order to extend the system to create longer melodies, it would be necessary to modify the RPI programs accordingly, as well as to extend the hardware of the beatbox controller and the melody controller, and to build a beatbox composer, and a Lego melody composer with more connectors, and an adequate multiplexing scheme.

## 3 Results and discussion

This interactive music demonstrator allows creating melodies with accompanying chords and percussion in an easy and intuitive way. Figure 19 and Fig. 20 show examples of use of the Lego-based music composition system developed. As the examples show, it is an easy-to-use system that allows the intuitive generation of melodies and rhythms in real time. Moreover, it should be noted that a simple snapshot of the Lego construction with the system designed represents the score of the melody and accompanying chords, and constitutes a simple way to save the creations; Figs. 19 and 20 illustrate this fact and the relation between the constructions built and the corresponding scores.

This system was presented at Transfiere 2020, obtaining great public success [8]. Transfiere 2020 is the main R + D + I meeting in Southern Europe to share scientific and technological knowledge, promote innovation and connect science and business. This edition was attended by about 4600 professionals from more than 1800 different entities, representing 30 different countries.

The comments received from the subjects playing with the system were all very positive. The system was tested by many of the attendees at Transfiere 2020, since it attracted a lot of attention as it was generating music continuously and its main means of interaction was the Lego pieces. All the participants testing this system found it appealing and easy to use; we think this is due the fact that the system designed and built is based on a very common and well-established construction system, and to the immediacy playing the music being created. People with musical knowledge required neither help nor previous explanations to use this music construction system, whereas non-musicians were able to construct their own melodies and music after a brief explanation.

Fig. 18 MIDI sequence with clock, rhythm, melody, and accompanying chord signals

| MIDI-MESSAGE (Hexadecimal) | EVENT |
|---|---|
| FA | Start |
| F8 | Timing Clock |
| F8 | Timing Clock |
| ⋮ | ⋮ |
| F8 | Timing Clock |
| B9 | Chan 10 Mode Change |
| 7B | All Notes Off |
| 00 | All Notes Off |
| 99 | Chan 10 Note On |
| 24 | Bass drum |
| 99 | Chan 10 Note On |
| 26 | Box |
| 99 | Chan 10 Note On |
| 31 | Charlie 1 |
| 99 | Chan 10 Note On |
| 2E | Charlie 2 |
| B0 | Chan 1 Mode Change |
| 7B | All Notes Off |
| 00 | All Notes Off |
| 90 | Chan 1 Note On |
| (00-7F) | Note 1 |
| 90 | Chan 1 Note On |
| (00-7F) | Note 2 |
| 90 | Chan 1 Note On |
| (00-7F) | Note 3 |
| 90 | Chan 1 Note On |
| (00-7F) | Note 4 |
| FB | Continue |
| F8 | Timing Clock |
| ⋮ | ⋮ |
| F8 | Timing Clock |
| ⋮ | ⋮ |
| adFB | Continue |
| ⋮ | ⋮ |
| FB | Continue |

(24 times for initial Timing Clocks; Rhythmic section; Melody and accompanying section; 24 times; 15 times)

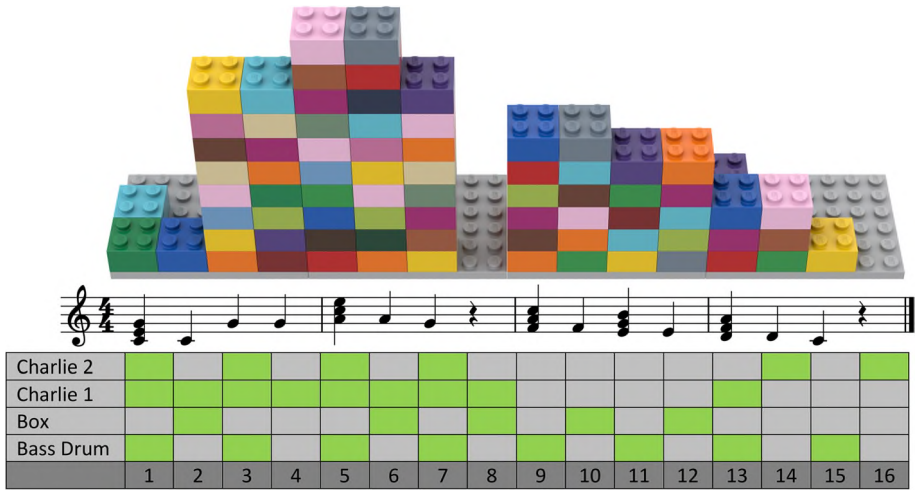

**Fig. 19** Excerpt of "Little start" constructed by using pitch-towers; its corresponding score and a proposed rhythmic pattern are also shown

The general feeling of all participants was that the system was useful for learning music in a very entertaining way. It is noteworthy that those who asked about the system spent some time constructing melodies, listening to them, making variations, etc. Also, it became clear that some training is required to achieve satisfactory results in terms of melody and rhythm working together, using the full system: the rhythmic pattern builder and the Lego melody and accompanying chords composer; however, the learning curve was not observed to be steep, on the contrary, it seemed really accessible for subjects with certain musical knowledge. On the other hand, the utilization of the composition subsystems separately was considered very simple, and lead to reasonably satisfactory results almost immediately.

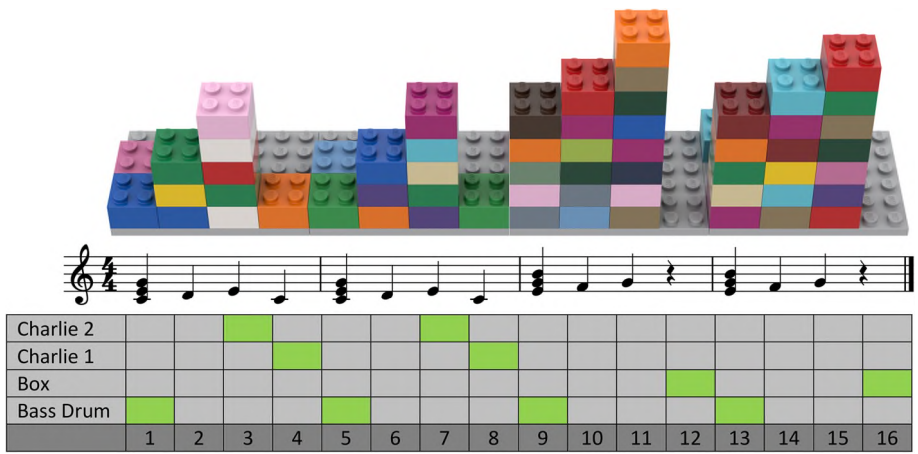

**Fig. 20** Excerpt of "FrÃ"re Jacques" constructed by using pitch-towers; its corresponding score and a proposed rhythmic pattern are shown

In general, the attendants that played with this music composition system recognized and emphasized its usefulness for entertaining young children while creating music. Moreover, its convenience for occupational therapy was also highlighted, specifically for elder people in need of both motor and cognitive rehabilitation or training and in therapies with autistic children.

There are some other advanced systems that use Lego pieces and sensors that got some social interest and relevance; those schemes are aimed at creating musical instruments in a way that encourages creativity and learning engineering concepts [10, 17]; even a microtonal guitar with Lego pieces has been designed [29]. However, a different perspective is considered in the design and construction of the system presented in this manuscript; in this case, our system, for the creation/construction of music, doesn't rely on the design of new instruments but on building music in a manipulative and simple way, linked to the natural interpretation of rhythm and height of pitch.

Music and building with Lego pieces are two well-differentiated tasks employed to stimulate different motor and intellectual functions of children [3, 4]. Considering this context, the system presented in this manuscript allows working both skills: to learn music and to improve fine motor skills simultaneously, and with high degree of interactivity and immediacy.

Music and Lego, separately, are also used as therapy to reduce post-operative pain in children [35]. With the system presented in this paper, music and Lego can be used together, and this kind of therapies could be improved by their simultaneous interrelation.

Also, for several years now, work has been done on new therapies for improving inclusion and social skills among children and youth with autism. Two commonly used elements to attain these objectives are Lego therapy [19] and music [36]. A system based on Lego, which allows creating music in an interactive and collaborative way, as the one presented in this paper, can be very useful for the treatment of children and youngsters with autism.

Regarding occupational therapy for the elderly, new apps and games are appearing [5] lately, as well as many works related to cognitive stimulation with music [6]. Precisely, this is one of the application environments highlighted by some users testing the system at Transfiere 2020 regarding the benefits that could be attained by using this system; the benefits specifically considered were related to psychomotricity and cognitive skills (Table 1 Appendix).

## 4 Conclusions

A system for building music melodies with a rhythmic base and chord accompaniment with Lego bricks using Raspberry Pi has been designed. The target audience is formed mainly of children, but the system developed is considered to be entertaining and beneficial for people of all ages.

The system consists of three main parts: one to synchronize the whole scheme (synchronism box), another one to set the rhythm (beatbox) and one to create the melody and accompanying chords (melody box). Each part of the system is controlled by a separate Raspberry Pi (RPI).

The modular structure designed allows the use of the complete system (rhythm and melody) or only a part of it (only rhythm or only melody). This gives the system a wide range of utilization possibilities.

The design of the system presented is radically different from other approaches found in the literature, which use Lego and music from the perspective of the construction of new instruments; our novel approach is linked to music representation in terms of rhythm and score, and the natural interpretation of the height of pitch. This approach makes the system designed easy to use, and together with is immediacy creating music, suitable for the rehabilitation of elderly people or for the cognitive stimulation of young children.

The use tests that have been carried out with the system in a real-world scenario have unveiled great interest, and discovered the usefulness of this system for learning and entertainment for both young and elder people. Some direct applications identified are: joint stimulation of children's motor and intellectual functions, therapy to reduce pain in the postoperative period of children, improvement of inclusion and social skills among children and youth with autism, and occupational therapy for the elderly.

With all this, it has been found that the design and implementation done are really satisfactory, although improvement and extensions could be considered, like the implementation of a simple hardware interface to set the BPM parameter, the possibility of cascading several subsystems like the one shown in this manuscript (however, note that this would require major redesign of some schemes, and further limitation of the BPM parameter), or the capacity to build with other musical figures.

**Acknowledgements** Alejandro Villena worked on several stages of the prototype for melody construction and synchronization. Joaquin Cáceres worked on the beatbox and synchronization.

**Funding** Funding for open access publishing: Universidad Málaga/CBUA This publication is part of the project PDC2021-120997-C33 funded by MCIN/AEI/10.13039/501100011033, and European Union "Next-GenerationEU/PRTR". This publication is part of the project PID2021-123207NB-I00 funded by MCIN/AEI/10.13039/501100011033/FEDER, UE. This work was done at Universidad de Málaga, Campus de Excelencia Internacional Andalucia Tech. Funding for open access charge: Universidad de Málaga/CBUA.

**Code availability** Code can be made available upon reasonable request.

## Declarations

**Conflict of interest/Competing interests** The authors have no financial or non-financial interests that are relevant to the content of this article.

# Appendix

**Table 1** Table of Acronyms

| Acronym | Meaning |
| --- | --- |
| BPM | Beats Per Minute |
| DOC | Digital Output Code |
| DIN | Deutsches Institut fur Normung e.V |
| GPIO | General Purpose Input Output |
| I2C | Inter-Integrated Circuit |
| MIDI | Musical Instrument Digital Interface |
| Mux | Multiplexer |
| RPI | Raspberry Pi |
| Rx | Receive |
| SPI | Serial Peripheral Interface |
| Tx | Transmit |
| UART | Universal Asynchronous Receiver-Transmitter |
| VST | Virtual Studio Technology |